\documentclass[twocolumn,nofootinbib,floatfix,superscriptaddress,aps,prd]{revtex4-1}

\usepackage[bookmarks=false]{hyperref}
\usepackage{graphicx}
\usepackage{amsmath}
\usepackage{xspace}
\usepackage{xcolor}
\usepackage{float}

\newcommand{\eq}[1]{Eq.~\eqref{eq:#1}}
\newcommand{\eqs}[2]{Eqs.~\eqref{eq:#1} and \eqref{eq:#2}}
\renewcommand{\sec}[1]{Sec.~\ref{sec:#1}}

\newcommand{\fig}[1]{Fig.~\ref{fig:#1}}

\newcommand{\app}[1]{Appendix~\ref{app:#1}}

\newcommand{\df}{\mathrm{d}}
\newcommand{\img}{\mathrm{i}}

\newcommand{\al}{\alpha}
\newcommand{\bt}{\beta}
\newcommand{\ga}{\gamma}
\newcommand{\Ga}{\Gamma}

\newcommand{\si}{\sigma}

\newcommand{\GeV}{\:\mathrm{GeV}}

\newcommand{\nn}{\nonumber}

\newcommand{\Pythia}{\textsc{Pythia}\xspace}

\allowdisplaybreaks[4]

\begin{document}


\title{Probing factorization violation with vector angularities}

\author{Pim Bijl}
\affiliation{Nikhef, 
	Science Park 105, 1098 XG, Amsterdam, The Netherlands\vspace{0.5ex}}

\author{Steven Niedenzu}
\affiliation{Nikhef, 
	Science Park 105, 1098 XG, Amsterdam, The Netherlands\vspace{0.5ex}}

\author{Wouter J.~Waalewijn\vspace{0.5ex}}
\affiliation{Nikhef,
	Science Park 105, 1098 XG, Amsterdam, The Netherlands\vspace{0.5ex}}
\affiliation{Institute for Theoretical Physics Amsterdam and Delta Institute for Theoretical Physics, University of Amsterdam, Science Park 904, 1098 XH Amsterdam, The Netherlands\vspace{0.5ex}}
	
\date{\today}

\begin{abstract}

Factorization underlies all predictions at the Large Hadron Collider, but has only been rigorously proven in a few cases. One of these cases is the Drell-Yan process, $pp \to Z/\gamma + X$, in the limit of small boson transverse momentum. We introduce a one-parameter family of observables, that we call vector angularities, of which the transverse momentum is a special case. This enables the study of factorization violation, with a smooth transition to the limit for which factorization has been established. Like the angularity event shapes,  vector angularities are a sum of transverse momenta weighted by rapidity, but crucially this is a vector sum rather than a sum of the magnitude of transverse momenta. We study these observables in  \Pythia, using the effect of multi-parton interactions (MPI) as a proxy factorization violation, finding a negligible effect in the case where factorization is established but sizable effects away from it. We also present a factorization formula for the cross section, that does not include factorization-violating contributions from Glauber gluons, and thus offers a baseline for studying factorization violation experimentally using vector angularities. Our predictions at next-to-leading logarithmic accuracy (NLL$'$) are in good in agreement with  \Pythia (not including MPI), and can be extended to higher order.
\end{abstract}

\maketitle

\section{Introduction}
\label{sec:intro}

All predictions for scattering processes at the Large Hadron Collider (LHC), rely on factorization. Factorization allows one to write the cross section $\sigma$ for a given process and measurement as some convolution of various ingredients. At the LHC this is essential to separate the  perturbatively-calculable partonic cross section $\hat \sigma$, from the nonperturbative dynamics of the incoming protons, described by parton distribution functions $f$. For example, for the Drell-Yan process $pp \to \ga/Z + X$, this takes the form 
\begin{align} \label{eq:fact_DY}
  \frac{\df \si}{\df Q\, \df Y} &= \sum_{i,j}  \int\! \df x_1\, f_i(x_1,\mu) \int\! \df x_2\, f_j(x_2,\mu)
  \nn \\ & \quad \times
   \frac{\df \hat \si_{ij}}{\df Q\, \df Y}(x_1,x_2,\mu)
\,.\end{align}
Here the sum on $i,j=g, u, \bar u, d, \dots $ runs over all parton flavors, whose momentum fractions $x_{1,2}$ are integrated over, $Q$ and $Y$ the invariant mass and rapidity of the vector boson, and $\mu$ the factorization scale.

When measurements significantly restrict the QCD radiation in the final state, the factorization becomes more involved. For example, if in the Drell-Yan process the transverse momentum of the boson is measured to be small compared to $Q$, this implies that hadronic radiation must be soft (low-energetic) or collinear (parallel to one of the incoming protons). In this case, the factorization involves transverse-momentum-dependent parton distributions.

While factorization is proven for the Drell-Yan process~\cite{Bodwin:1984hc,Collins:1984kg,Collins:1988ig}, it is used for generic LHC processes. However, a crucial and nontrivial step to establish factorization involves showing that the Glauber region (or Glauber modes~\cite{Donoghue:2009cq,Fleming:2014rea,Rothstein:2016bsq} in Soft-Collinear Effective Theory (SCET)~\cite{Bauer:2000yr,Bauer:2001ct,Bauer:2001yt,Bauer:2002nz,Beneke:2002ph}) does not give a non-trivial contribution.\footnote{We write ``non-trivial", since Ref.~\cite{Rothstein:2016bsq} shows that certain ``Cheshire" Glauber contributions can simply be accounted for by using the proper orientation of Wilson lines.} There has been progress in understanding the origin of this factorization violation, finding e.g.~that for single-scale observables such contributions necessarily involve a Lipatov vertex~\cite{Schwartz:2018obd}. A concrete example of factorization violation was presented in Ref.~\cite{Zeng:2015iba}. For other work on factorization violation, see e.g.~Refs.~\cite{Collins:2007nk,Mulders:2011zt,Catani:2011st,Forshaw:2012bi}.

\begin{figure*}[t] 
\centering
    \includegraphics[width=\columnwidth]{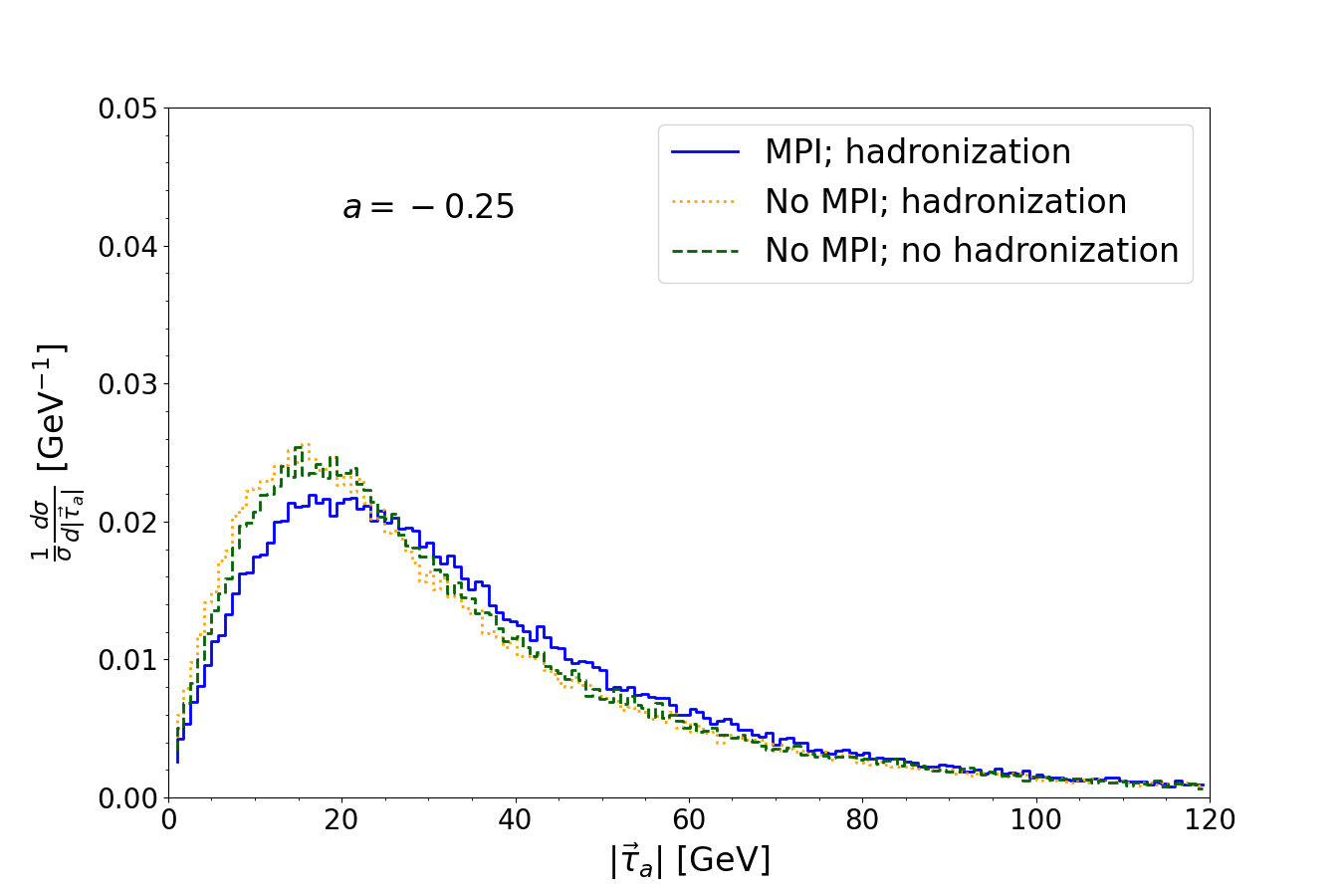}
    \includegraphics[width=\columnwidth]{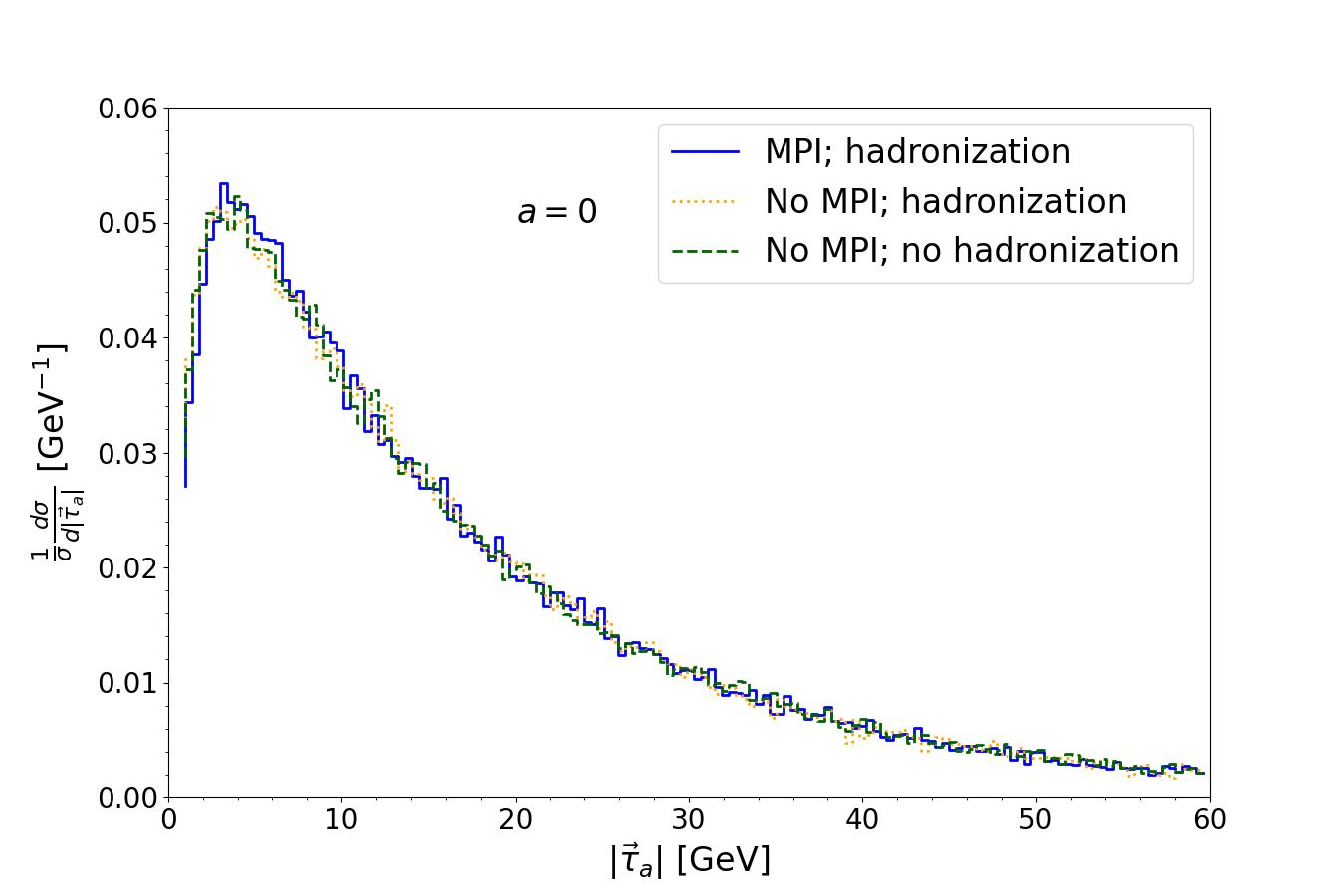}
    \includegraphics[width=\columnwidth]{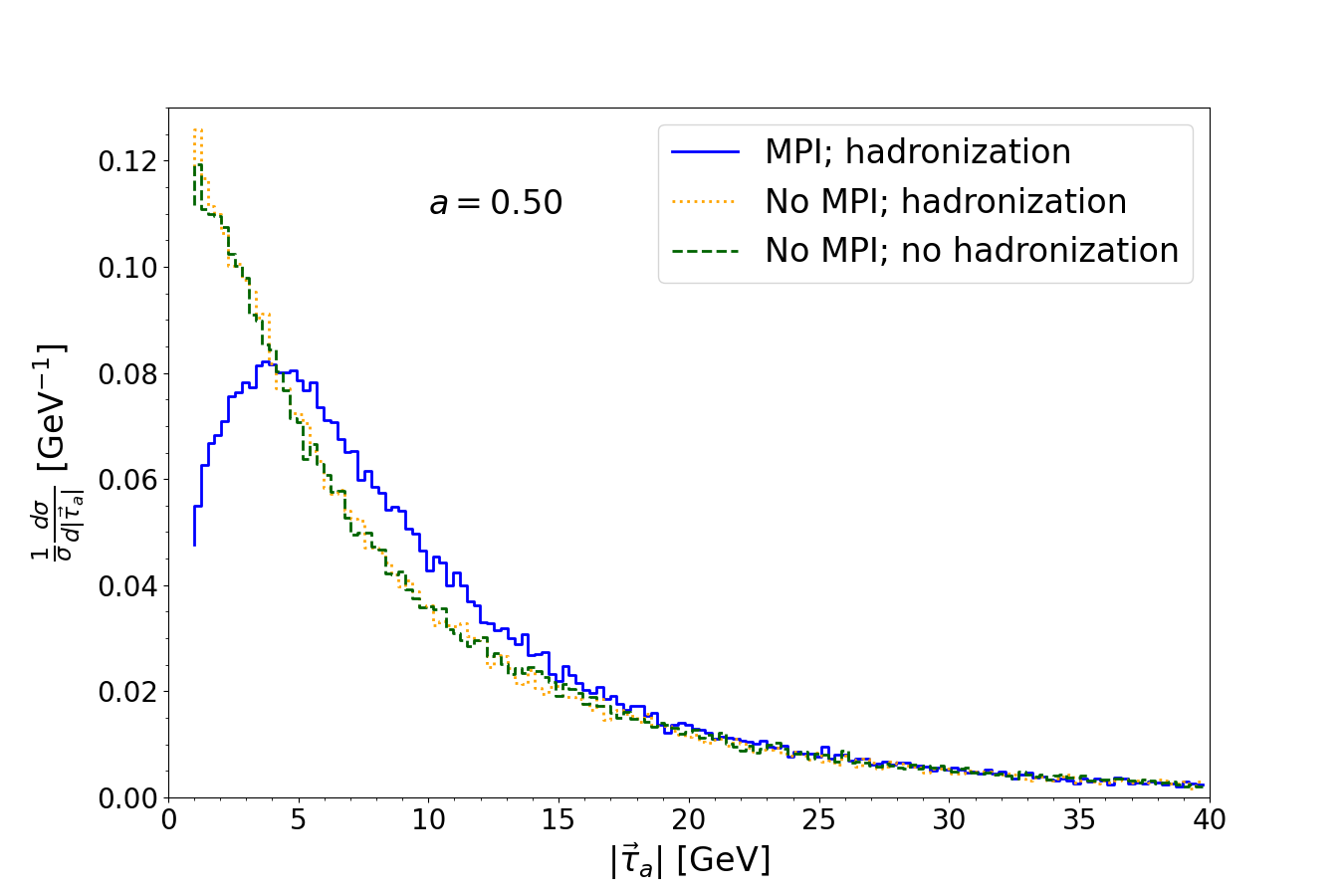}
    \includegraphics[width=\columnwidth]{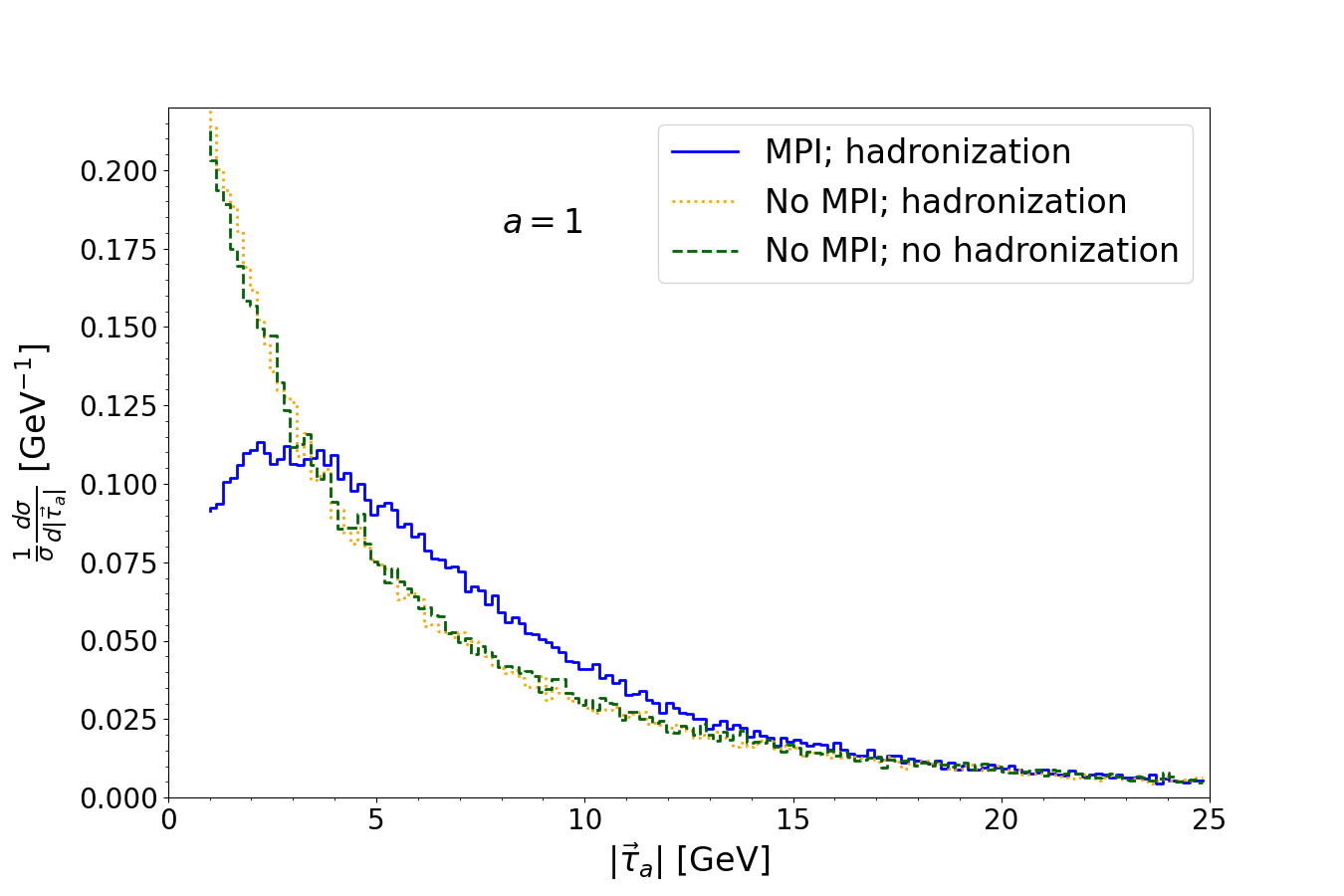}   \vspace{-2ex}
    \caption{Normalized cross sections differential in the vector angularity $|\vec \tau_a|$, for $a=-0.25$ (top left), 0 (top right), 0.5 (bottom left), 1 (bottom right), obtained from \Pythia as described in \sec{pythia}. Shown are the results without MPI and no hadronization (green dashed), without MPI but hadronization turned on (orange dotted) and with both MPI and hadronization (blue).\label{fig:pythia}}
\end{figure*}

In this paper we introduce a family of observables, which we call vector angularities. Though we focus on the Drell-Yan process, these observables can be applied to other processes in which a color singlet is produced. The vector angularities are defined as\footnote{Alternatively, one can include the boost of the $\ga/Z$ in the definition of the vector angularities, $y_i \to y_i - Y$. This simplifies the factorization in \eqs{factorizedcross}{final_cross}, eliminating the explicit $Y$ dependence in the convolution of $\vec{\tau}_a$~\cite{Stewart:2009yx,Gangal:2014qda}.}
\begin{equation}  \label{eq:vector_angularities}
 \vec{\tau}_a = \sum_i \vec{k}_{\perp,i} e^{-a|y_i|},
\end{equation}
where the sum on $i$ runs over the hadronic final state, and the transverse momentum $\vec{k}_{\perp, i}$ of particle $i$ is weighed by its rapidity $y_i$, in accordance to the choice of the parameter $a$. This is similar to the angularity event shape~\cite{Berger:2003iw}, which has been extended to Drell-Yan~\cite{Stewart:2009yx}, deep-inelastic scattering~\cite{Kang:2021yaf} and jets~\cite{Almeida:2008yp}. A crucial difference is that we take the \emph{vector} sum of transverse momenta. 

For the special case of $\vec \tau_0$, this corresponds to the transverse momentum of the hadronic final state, and thus the transverse momentum of the boson, by momentum conservation. In this case it has been shown that the factorization-violating contributions from the Glauber region can indeed be ignored. Our family of observables therefore allows one to explore possible factorization-violating effects in a way that lets one smoothly turn them off as $a \to 0$.

The effect of Glauber gluons has been connected to multiple-parton interactions (MPI)~\cite{Gaunt:2014ska}, i.e.~multiple partonic collisions between the same pair of colliding protons. As in Ref.~\cite{Kang:2018agv}, we therefore study the effect of MPI in \Pythia~\cite{Sjostrand:2014zea} on our vector angularities to get a first impression of possible factorization violation. 

We will also present a factorization formula for the Drell-Yan cross section differential in $\vec \tau_a$, assuming the \emph{absence} of factorization-violating effects. Interestingly, this vector-type observable does not involve rapidity divergences, and it is another example where resummation needs to be carried out in the conjugate space~\cite{Bhattacharya:2023qet}. We obtain resummed predictions at next-to-leading logarithmic (NLL$'$) accuracy, which are in agreement with \Pythia (without MPI). Though the accuracy of these predictions is limited, they can in principle be substantially improved by including higher-order corrections. Indeed, for $\vec \tau_0$ results for Drell-Yan have even been obtained at next-to-next-to-next-to-next-to-leading logarithmic (N$^4$LL) accuracy!~\cite{Neumann:2022lft,Moos:2023yfa,Camarda:2023dqn} Such resummed calculations would provide a baseline for studying factorization violation experimentally, using vector angularities. 

The outline of this paper is as follows: In \sec{pythia} we show numerical results from \Pythia for the vector angularities, to explore their sensitivity to factorization violation using its MPI model. Our factorization formula and resummed results are presented in \sec{calculation}, with the perturbative ingredients relegated to \app{ingredients}. We conclude in \sec{conclusions}

\section{Pythia and Underlying Event}
\label{sec:pythia}

In this section we study the effect of MPI in the \Pythia Monte Carlo event generator to assess the sensitivity of vector angularities $\vec \tau_a$ to factorization violation. In particular, we are interested in studying the dependence on $a$, knowing that for $a=0$ factorization violation effects should be absent.

We simulate proton-proton collisions in \Pythia with a center-of-mass energy of 13 TeV. One parton from each colliding proton engages in a hard scattering process to produce a $Z$ boson/photon (the Drell-Yan process). For definiteness, we set the subsequent decay of this boson to an electron-positron pair. There can be additional interactions between the remaining partons in the protons. These MPI lead to additional radiation, i.e.~on top of that emitted in the production of a $Z$ boson/photon. We will furthermore always include initial-state radiation, and explore the effect of hadronization by turning it on/off. For the PDFs we use MSTW2008 at next-to-leading order (NLO)~\cite{Martin:2009bu} with $\alpha_s(M_Z) = 0.12$, which we also employ in our analytic calculations in \sec{calculation}. 

For each simulated event, the vector angularity $\vec \tau_a$ in \eq{vector_angularities} is calculated by summing over the contribution from each final-state particle, except for the decay products of the boson. Due to the fact that other processes besides the boson decay can also create electrons or positrons, we need to select the proper final-state particles to calculate the vector angularities. Events are selected based on the requirement that exactly one electron and one positron have a transverse momentum of $k_{\perp} \geq 2.5$ GeV, which we assume to be from the boson decay and are not included in the calculation of $\vec \tau_a$. Only 2 percent of the events are discarded due to this cut. 

The simulations are run for three scenarios: First we have MPI and hadronization turned off, which can be compared to our analytical calculation. Second, MPI and hadronization are both on, which should be representative of measurements of the LHC, though a full study would require including backgrounds and detector simulations. Third, hadronization is kept on but MPI is turned off to assess the size of factorization-violating effects.
For each simulation scenario, different values of the parameter $a$ were considered, ranging from $a = -0.25$ to $a = 1$. This range was chosen to explore the vicinity of $a=0$, where the effect of MPI is expected to be negligible. Furthermore, for $a \leq -1$, we would run into problems of IR safety, while for very large values of $a$ only radiation at extremely central rapidities would be probed. We generate 100.000 events and calculate the vector angularities for each value of $a$. Because the angle of $\vec \tau_a$ is irrelevant, we show results for $|\vec \tau_a|$.

The results are shown in \fig{pythia}, where we choose $a = -0.25, 0, 0.5$, and 1 as representative values. First we note by comparing the curves with/without hadronization, that its effect is rather mild. (Though we also explored more negative values of $a$, such as $a=-1$, where hadronization effects are large.) For $a = 0$, MPI clearly have very little effect on the vector angularities. This is in agreement with the expectation for $\vec \tau_0$, for which factorization-violating effects from Glauber gluons are known to be absent. For the other values of $a$, there is a more pronounced difference between the distribution that includes MPI and those that do not, and this difference increases as $a$ is further from 0. These same conclusions are reached if \textsc{Herwig}~\cite{Bellm:2017bvx} is used: though the individual curves are slightly different, the effect of  MPI is very similar.\footnote{We thank A.~Papaefstathiou for help with verifying this.}
 Based on this we conclude that the vector angularities are indeed sensitive to MPI, suggesting their use as a probe of factorization violation. 
 
\ \\

\section{Resummed calculation}
\label{sec:calculation}

\subsection{Factorization}

Using the framework of SCET, we factorize the cross section for small $|\vec \tau_a|$ into the following ingredients
\begin{widetext}
\begin{align} \label{eq:factorizedcross}
   \frac{\df \si}{\df Q\, \df Y\, \df^2 \vec \tau_a} &= \sum_q \si_{0,q} H(Q^2,\mu)
   \int\! \df^2 \vec \tau_{a,1}'\, B_q(\vec \tau_{a,1}',x_1,\mu)
   \int\! \df^2 \vec \tau_{a,2}'\, B_{\bar q}(\vec \tau_{a,2}',x_2,\mu)\,
   S\Bigl(\vec \tau_a - \frac{\vec \tau_{a,1}'}{(Qe^Y)^a} - \frac{\vec \tau_{a,2}'}{(Qe^{-Y})^a},\mu\Bigr)
   \nn \\ &
   = \sum_q \si_{0,q} H(Q^2,\mu)
   \int \! \frac{\df^2 \vec b_\perp}{(2 \pi)^2} e^{-\img \vec \tau_a \cdot \vec b_\perp}
    \tilde B_q\Bigl(\frac{\vec b_\perp}{(Qe^Y)^a},x_1,\mu\Bigr)
   \tilde B_{\bar q}\Bigl(\frac{\vec b_\perp}{(Qe^{-Y})^a},x_2,\mu\Bigr)\,
   \tilde S(\vec b_\perp,\mu)
\,,\end{align}
\end{widetext}
where we obtained the second line by performing a Fourier transform. Our convention for the Fourier transform is given in \eq{fourier}, and we include a tilde to indicate that the functions have been transformed.
The sum on $q=u, \bar u, d, \dots $ runs over all (anti-)quark flavors, and the momentum fractions $x_1 = Qe^Y/E_{\rm{cm}}$, $x_2 = Qe^{-Y}/E_{\rm{cm}}$ are fixed by the invariant mass $Q$ and rapidity $Y$ of the vector boson, and the center-of-mass energy $E_{\rm{cm}}$ of the collision. 
    
The Born cross section is given by
\begin{align}
        \si_{0,q} & = \frac{8\pi \al_{\rm{em}}^2}{3N_c Q  E_{cm}^2}
       \\ & \quad \times
       \biggl[ Q_q^2 \!+\! \frac{(v_q^2 \!+\! a_q^2) (v_\ell^2\!+\!a_\ell^2) \!-\! 2 Q_q v_q v_\ell (1\!-\!m_Z^2/Q^2)}
{(1\!-\!m_Z^2/Q^2)^2 \!+\! m_Z^2 \Gamma_Z^2/Q^4} \biggr],
\nn\end{align}
where $\al_{\text{em}}$ is the electromagnetic coupling, $N_c=3$ is the number of colors, $Q_q$ is the electric charge of the quark, $v_{\ell,q}$ and $a_{\ell,q}$ are the standard vector and axial couplings of the leptons and quarks, and $m_Z$ and $\Gamma_Z$ are the mass and width of the $Z$ boson. 
    
    The hard function $H$ describes the short-distance collision of an incoming quark and anti-quark that produce the $Z/\gamma$, and includes virtual hard corrections. Real hard radiation is not possible for small $|\vec \tau_a|$, and so the hard function is independent of the vector angularity measurement. $B_q$ and $B_{\bar q}$ are the beam functions of the incoming (anti-)quark, which include the PDFs and initial-state radiation coming from the extracted partons~\cite{Stewart:2009yx}. At next-to-leading order, this includes a contribution from the gluon PDF, where the extracted gluon splits into a quark and anti-quark pair, with one of them entering the hard interaction and the other going into the final state. The contribution of soft radiation to the measurement, emitted by the incoming partons, is encoded in the soft function $S$. 
    
The hard function was calculated a long time ago~\cite{Manohar:2003vb,Bauer:2003di}. We have calculated the one-loop beam and soft function, and present our results in \app{pert}. As these functions contain at most one real emission, there is only a small but subtle difference~\footnote{The measurement of the scalar angularity involves the $(d\!-\!2)$-dimensional transverse momentum, while the vector angularity  involves the 2-dimensional part. For details on the treatment of transverse momenta in dimensional regularization see e.g.~\cite{Jain:2011iu,Lubbert:2016rku}. This affects the $\pi^2$-term in the one-loop soft and beam functions, though not in their sum.} between a vector or scalar sum, allowing us to validate our results with the known angularity soft~\cite{Hornig:2009vb} and beam function~\cite{Zhu:2021xjn}. The difference between vector and scalar sum does lead to distinct features for the resummation, since this encodes the dominant effect of multiple emissions.

\subsection{Resummation}
\label{sec:resummation}

To perform the resummation, we evaluate the hard, beam and soft functions at their natural scale and use the renormalization group to evolve them to a common scale. In the case of transverse momentum resummation,  performing the resummation directly in momentum space is challenging (see e.g.~Refs.~\cite{Frixione:1998dw,Ebert:2016gcn}), so we will also switch to Fourier space for $\vec \tau_a$.

The resummed Drell-Yan cross section differential in $Q$, $Y$ and $|\vec \tau_a|$ is given by
\begin{align} \label{eq:final_cross}
    &\frac{\df \si}{\df Q\, \df Y\, \df |\vec \tau_a|} 
     \\
    & \quad = \sum_q \si_{0,q} H(Q^2,\mu_H)\,
     \int_0^\infty \! \df b_\perp\, b_\perp  |\vec \tau_a|\, J_0\bigl(b_\perp |\vec \tau_a|\bigr)
    \nn \\ 
    & \qquad \times
    \tilde B_q\Bigl(\frac{b_\perp^*}{(Qe^Y)^a},x_1,\mu_B\Bigr)\,
   \tilde B_{\bar q}\Bigl(\frac{b_\perp^*}{(Qe^{-Y})^a},x_2,\mu_B\Bigr)\,    
    \nn \\
   & \qquad \times
    \tilde S(b_\perp^*,\mu_S)\,
    U_H(Q^2,\mu_H,\mu_B)\,    
   U_S(b_\perp^*, \mu_S, \mu_B, a) \,.
\nn \end{align}
We have now written the cross section differential in $|\vec \tau_a|$, and indicated that the soft and beam functions depend on $b_\perp \equiv |\vec b_\perp|$ and not the angle of $\vec b_\perp$ (due to azimuthal symmetry).
Compared to \eq{factorizedcross}, we have included the evolution kernels $U_H$ and $U_S$ of the hard and soft function, that we use to evolve them from their natural scale $\mu_H$ and $\mu_S$ to the beam scale $\mu_B$. The expressions for the renormalization group equations and evolution kernels are given in \app{resum}, and the natural scales $\mu_H, \mu_B$, and $\mu_S$ are discussed below. Finally, the star in $b_\perp^*$ indicates a prescription to avoid the Landau pole, which also enters through the scales $\mu_B$ and $\mu_S$, see \eq{bstar}.

The natural scales of the hard, beam, and soft functions are those for which they do not contain large logarithms. As the hard function contains logarithms of  $Q^2/\mu_H^2$, see \eq{hard}, the natural scale is $\mu_H = Q$. In the soft and beam functions the large logarithms are $L_b$ and $L_b'$ in \eqs{Lb}{Lbp}. We simply choose $\mu_S$ and $\mu_B$ such that $L_b = 0$ and $L_b^{'} = 0$, respectively. If we strictly follow this procedure, $\mu_B$ would depend on $Y$, so we instead choose to use the scale obtained in this manner for $Y=0$. 
The natural scales are then given by 
\begin{align}
        \mu_H &= Q, \nn \\
        \mu_S &= \frac{2e^{-\gamma_E}}{ b_\perp^{*}}, \nn \\
        \mu_B &= Q^{\frac{a}{1+a}} \Bigl(\frac{2e^{-\gamma_E}}{b_\perp^{*}}\Bigr)^{\frac{1}{1+a}}.
\end{align}
 
The uncertainties on the cross section are determined by varying the hard, beam and soft scales up and down. For the hard scale this is a factor of two, i.e.~we take $\mu_H=Q$ for the central curve and consider $\mu_H=Q/2$ and $\mu_H=2 Q$ to estimate the perturbative uncertainty. For the soft scale we also take a factor of 2 but simultaneously vary $\mu_B$ in order to maintain $\mu_B^{1+a} = \mu_H^{a} \mu_S$. The $\mu_S$ variation dominates our scale uncertainty, and we symmetrize the resulting uncertainty band. 

\begin{figure*}[t] 
\centering
    \includegraphics[width=\columnwidth]{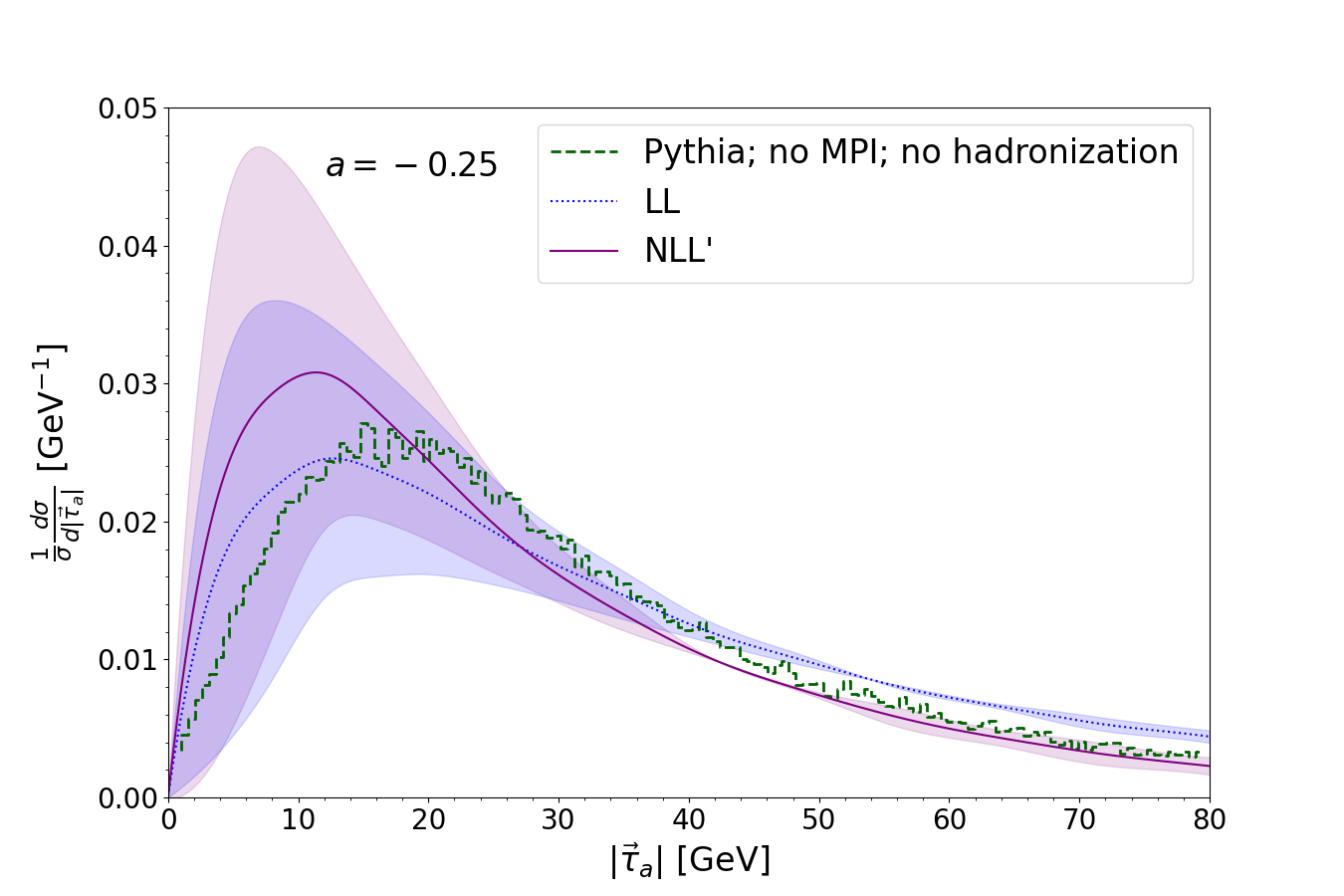}
    \includegraphics[width=\columnwidth]{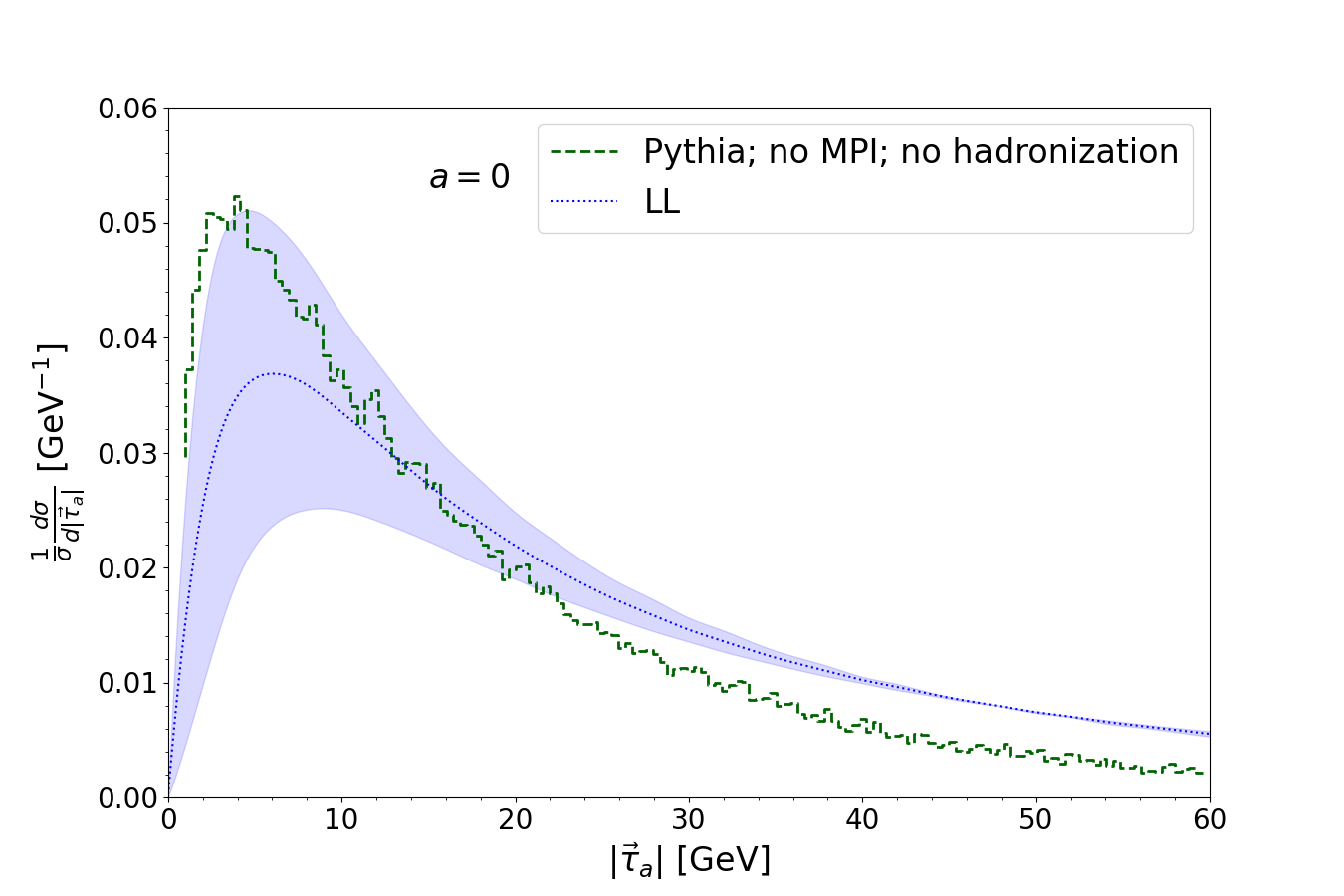}
    \includegraphics[width=\columnwidth]{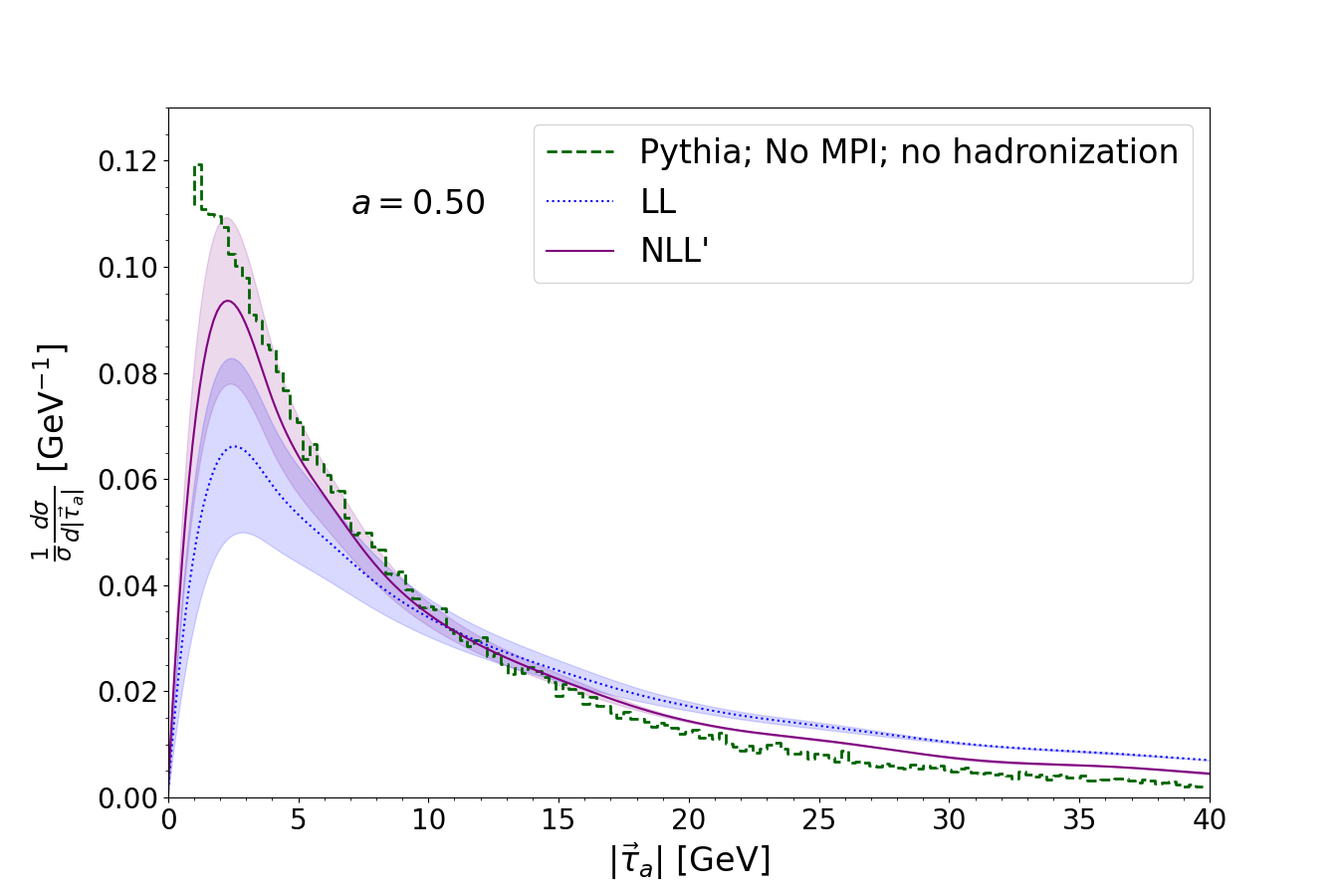}
    \includegraphics[width=\columnwidth]{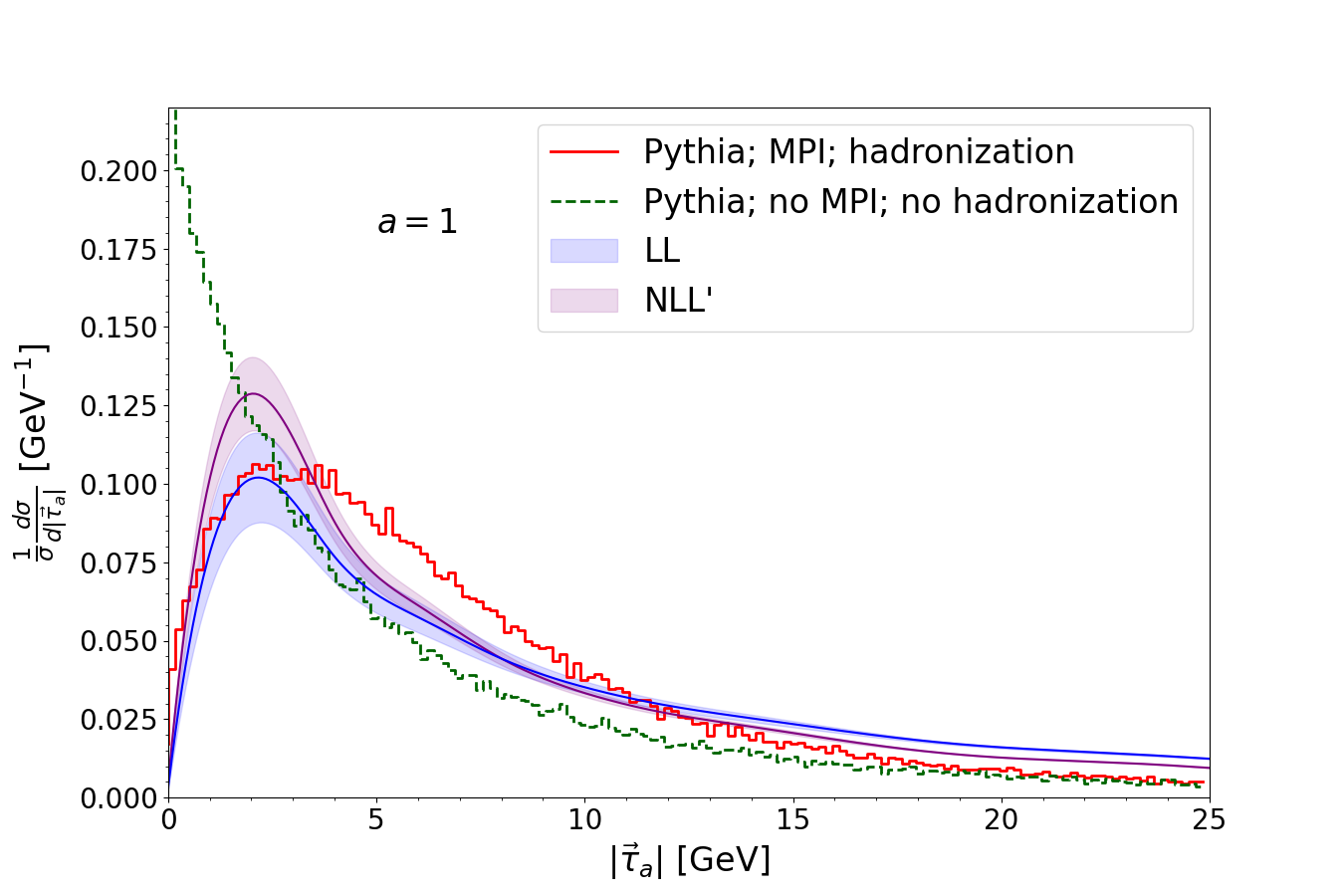} \vspace{-2ex}
    \caption{Normalized cross sections differential in the vector angularity $|\vec \tau_a|$, for $a=-0.25$ (top left), 0 (top right), 0.5 (bottom left), 1 (bottom right), obtain using our resummed calculation at LL (blue dotted) and NLL$'$ (purple). The bands indicate the perturbative uncertainty, estimated using the scale variations in \sec{resummation}. The \Pythia result without MPI and without hadronization (green dashed) from \fig{pythia}, is shown for comparison.\label{fig:results}}
\end{figure*}

To avoid the Landau singularity, we regularize the nonperturbative region at large $b_\perp$ in \eq{final_cross} by using a ``$b$-star" prescription~\cite{Collins:1984kg}, 
\begin{equation} \label{eq:bstar}
b^*_\perp=\frac{b_\perp}{\sqrt{1+ b_\perp^2/b_{\rm max}^2}}
\,.\end{equation}
This ensures that $b^*_\perp \to b_{\rm max}$ when $b_\perp \to \infty$.
We determine an appropriate value for $b_{\rm max}$ by requiring that the scales $\mu_H, \mu_B$, and $\mu_S$ stay above the minimum value $\mu_0=0.8 \GeV$. Since the beam scale depends on the parameter $a$, the expressions for $b_{\rm max}$ will depend on the value of $a$ that is considered: For $a \geq 0$, $\mu_S$ is the smallest scale, while for $a < 0$, $\mu_B$ is the smallest scale. This leads to
\begin{align}
        b_{\rm max}(a) &= 
          \frac{2 e^{-\gamma_E}}{\mu_0} \Bigl(\frac{Q}{\mu_0}\Bigr)^{\text{min}(a,0)}
\,.\end{align}

For our normalized distributions, the integration over $Q$ is numerically irrelevant (percent-level effect). Performing the $Y$ integral changes the shape of the $\vec \tau_a$ distribution by less than 10\% compared to using $Y=0$, and is extrapolated from the leading logarithmic (LL) cross section using,
\begin{align}
  \int\! \df Y\, \frac{\df \si_{{\rm NLL}'}}{\df Y\, \df^2 \vec \tau_a} &=
  \biggl[\frac{\df \si_{{\rm NLL}'}}{\df Y\, \df^2 \vec \tau_a} \biggl/ \frac{\df \si_{\rm LL}}{\df Y\, \df^2 \vec \tau_a} \biggr]_{Y=0}  
  \nn \\ & \quad \times
  \int\! \df Y\, \frac{\df \si_{\rm LL}}{\df Y\, \df^2 \vec \tau_a}
\,.\end{align}

Finally, we note that for large $b$, $b_\perp^* \to b_{\rm max}$ and the cross section in $b$ approaches a constant. A constant in $b$-space transforms to a delta function in $\vec \tau_a$, and we therefore subtract of this contribution to improve the numerical stability. Alternatively, one can include a nonperturbative model $\exp(-\Lambda \vec b_\perp^{\,2})$ to suppress this region, which for $\Lambda = 0.35$ GeV yields the same within a few percent.

\subsection{Numerical results}
\label{sec:results}

We have implemented our resummation in \eq{final_cross} at LL and NLL$'$ order. The former only involves the tree-level hard, beam and soft function as well as the lowest order cusp anomalous dimension $\Ga_0$ and running coupling ($\beta_0$). At NLL$'$ we include all ingredients in \app{ingredients}, and consistently expand the cross section, e.g.~dropping cross terms involving a one-loop beam and one-loop soft function. We do not include the matching to the NLO cross section, so our results become less reliable for large values of $|\vec \tau_a|$. For this reason, we normalize the distribution on the plotted interval.

Our results  for $a = -0.25, 0, 0.5, 1$ at LL and NLL$'$ are shown in \fig{results}. The bands indicate the perturbative uncertainty and are obtained by scale variations, as discussed in \sec{resummation}. We apply the same factor for the scale variations as was used to normalize the central curve, instead of separately normalizing each of the scale variations. This makes our uncertainty estimate more conservative. Formally our expressions diverge at $a=0$, so in this case we take the limit numerically by setting $a=0.01$~\footnote{The divergence at $a=0$ in the one-loop soft- and beam functions requires a rapidity regulator, see e.g.~Refs.~\cite{Collins:2011zzd,Echevarria:2011epo,Chiu:2012ir}. We have checked that the $1/a$ poles cancel when summing the one-loop contributions from the soft- and beam functions, and that the result in the $a\to 0 $ limit agrees with the known result for transverse momentum factorization~\cite{Becher:2010tm,Ritzmann:2014mka}. The soft anomalous dimension ``converts" into the rapidity anomalous dimension in this limit, as discussed for the angularity event shape in $e^+e^-$ collisions with respect to a \emph{recoil-free} axis~\cite{Larkoski:2014uqa,Bell:2018vaa}.}. This yields a reasonable result at LL, but the NLL$'$ curve is unstable due to large cancellations between the one-loop soft and beam functions and therefore omitted. For the other values of $a$, the LL and NLL$'$ uncertainty bands tend to overlap, except for larger values of $|\vec \tau_a|$. There, our results are anyway less reliable because our factorization formula does not account for power corrections of $\mathcal{O}(|\vec \tau_a|^2/Q^2)$, which could be remedied by matching to the NLO cross section. For $a=0.5$ and 1, the uncertainties at NLL$'$ are smaller than at LL, indicating convergence. For $a = -0.25$, the (relative) uncertainties at LL and NLL$'$ are very similar. This is in line with the results in Ref.~\cite{Procura:2018zpn} for angularities in jets, see their Fig.~7, where their $\beta$ corresponds to $a+1$. 

Finally, these plots also include the \Pythia results at parton level, without MPI, which are in agreement with our calculation. Since \Pythia contains NLL$'$ ingredients it is not surprising they are closer to NLL$'$ than LL.
The agreement is not as good for $a=1$, indeed for small values of the vector angularity our result is closer to  \Pythia with MPI. (We have included the corresponding curve only in this case, to keep the other plots clear.) This emphasizes that to extract factorization-violating effects experimentally, it is important to have a good baseline prediction without these effects.
 We note that our calculation is of course only a first step, and expect substantial improvement at higher orders in perturbation theory.  In particular, the soft function needed at NNLL$'$+NNLO can be obtained using \textsc{SoftSERVE}~\cite{Bell:2018vaa,Bell:2018oqa,Bell:2020yzz}, and there are ongoing efforts to automate the beam function calculation at this order~\cite{Bell:2021dpb}.

\section{Conclusions}
\label{sec:conclusions}

In this work we proposed a new one-parameter family of hadron collider observables, called vector angularities, that can be used to study the effects of factorization violation. When the parameter $a=0$, factorization has been established and these effects are absent. Exploring factorization violation, using the MPI model of \Pythia as a proxy, we found agreement with the absence of these effects for  $a=0$, while they grow for values of $a$ away from 0. We also explored the effect of hadronization in \Pythia, which is tiny in comparison to MPI. 

We then presented a factorization formula for the vector-angularity cross section, assuming the absence of factorization violation. This would provide a baseline for studying these effects at the LHC. Our numerical results at LL and NLL$'$ were in agreement with \Pythia (without MPI), but still have large uncertainties. Calculating higher orders will certainly reduce this, and due to the development of automated tools, NNLL$'$+NNLO should soon be within reach. Indeed, for the special case of $a=0$, N$^4$LL results have already been obtained. Finally, it would also be very interesting to attempt a direct calculation of the contribution from Glauber gluon exchanges, using the formalism of Ref.~\cite{Rothstein:2016bsq}.
 
\begin{acknowledgments}
We thank D.~Neill for discussions.
This work is supported by the D-ITP consortium, a program of NWO that is funded by the Dutch Ministry of Education, Culture and Science (OCW).
\end{acknowledgments}

\appendix

\section{}
\label{app:ingredients}

We define our Fourier transformation as follows:
\begin{align} \label{eq:fourier}
  \tilde f(\vec b_\perp) = \int \df^2 \vec \tau_a\, e^{\img \vec \tau_a \cdot \vec b_\perp} f(\vec \tau_a)
.\end{align}

We now present the expressions for the ingredients in the factorization up to next-to-leading order, as well as the renormalization group equations and anomalous dimensions needed for NLL$'$ resummation. The $\overline{\text{MS}}$ scheme is employed through out.

\subsection{Perturbative ingredients}
\label{app:pert}

The renormalized hard function is given by
\begin{align} \label{eq:hard}
    H(Q^2,\mu)
    &= 1+ \frac{\alpha_s C_F}{2 \pi} \biggl(-\ln^2{\Bigl(\frac{Q^2}{\mu^2}\Bigr)}  + 3\ln{\Bigl(\frac{Q^2}{\mu^2} \Bigr)}
     \nn \\ & \quad
    -8 + \frac{7\pi^2}{6} \biggr) + \mathcal{O}(\alpha_s^2)
\,.\end{align}
The renormalized soft function is given by
\begin{align}
    \Tilde{S}(\Vec{b},\mu)&= 1 + \frac{\alpha_s C_F}{2\pi} \frac{1}{a} \Bigl(-L_b^2-\frac{\pi^2}{6}
    \Bigr) + \mathcal{O}(\alpha_s^2)\
\,,\end{align}
where 
\begin{align} \label{eq:Lb}
L_b \equiv \ln \bigl(\vec b_\perp^{\,2} \mu^2 e^{2\ga_E}/4\bigr)
\,.
\end{align}
The renormalized beam functions are matched onto PDFs using
\begin{equation}
    \tilde B_q(\vec b_\perp',x,\mu) = \sum_j \int \frac{\df x'}{x'} \tilde {\mathcal{I}}_{qj}(\vec b_\perp',x',\mu) f_j\left(\frac{x}{x'},\mu\right)
\end{equation}
with matching coefficients
\begin{widetext}
\begin{align}
    \tilde {\mathcal{I}}_{qq}(\vec b_\perp',x,\mu) & = \delta(1-x) +
    \frac{\alpha_s C_F}{2 \pi}\biggl[\frac{1}{2a(1+a)}(L_b')^2 \delta(1-x) + \frac{1}{1+a} L_b' (-2 \mathcal{L}_0(1-x)+x+1) 
    \nn \\
    & \qquad + \frac{4a}{1+a}\mathcal{L}_1(1-x) + \frac{1-a}{a}\frac{\pi^2}{12}\delta(1-x) - \frac{2a}{1+a}(1+x) \ln{(1-x)
     -\frac{2a}{1+a} \frac{1+x^2}{1-x} \ln x
    -x+1} \biggr] +\mathcal{O}(\alpha_s^2) \,, 
    \nn \\
    \tilde {\mathcal{I}}_{qg}(\vec b_\perp',x,\mu) & = \frac{\alpha_s T_F}{2 \pi}\left[-\frac{1}{1+a} L_b'(2x^2 - 2x +1) - (2x^2 - 2x +1)\Bigl(-\frac{2a}{1+a}(\ln{(1-x)}-\ln x) + 1 \Bigr) +1 \right] \!+\! \mathcal{O}(\alpha_s^2)
\,.\end{align}
\end{widetext}
Here 
\begin{align} \label{eq:Lbp}
L_b' \equiv \ln \bigl[(\vec b_\perp')^2 \mu^{2+2a} e^{2\ga_E}/4\bigr] = L_b  - 2a \ln(p^-/\mu)
\,,\end{align}
with $p^- = Q e^{\pm Y}$ depending on the beam. 

\subsection{Renormalization group evolution}
\label{app:resum}

The one-loop anomalous dimensions of the hard, beam and soft function are 
\begin{align}
    \gamma_H^{(1)} &= \frac{\alpha_s C_F}{\pi}\biggl[2 \ln{\Bigl(\frac{Q^2}{\mu^2} \Bigr)} -3 \biggr]\,, \nn \\ 
    \gamma_S^{(1)} &= \frac{\alpha_s C_F}{\pi} \Bigl( \frac{-2}{a} L_b  \Bigr)\,, \nn \\
    \gamma_B^{(1)} &= \frac{\alpha_s C_F}{\pi} \Bigl(\frac{1}{a}L_b'+\frac{3}{2} \Bigr)\,.
\end{align}
As required by consistency of the factorization, 
\begin{equation}
\ga_H + \ga_S + 2 \ga_B = 0
\,,
\end{equation}
where we use that $p_1^- p_2^+ = Q e^Y Q e^{-Y} = Q^2$.
To achieve NLL$'$ accuracy, we need to include the two-loop cusp anomalous dimension, $\Gamma_1$, given in \eq{gabeta}.

We carry out the resummation by evolving the hard and soft function to beam scale. This requires solving the differential equation 
\begin{align}
  \frac{\df}{\df \ln \mu}\, F = \gamma_F\, F
\,,\end{align}
which in general involves a convolution between $\ga_F$ and $F$, but is multiplicative for the Fourier conjugate variables $\vec b_\perp$ and $\vec b_\perp'$. The evolution kernels for evolving the hard and soft function from a scale $\mu_0$ to $\mu$ are given by:
\begin{widetext}
\begin{align} \label{eq:evokernel}
    U_H(Q^2,\mu_0,\mu) 
&= \exp\biggl[-4 K_\Gamma + \frac{6 C_F}{\bt_0} \ln r \biggr]\biggl(\frac{Q^2}{\mu_0^2}\biggr)^{2\eta_\Ga}\,,
   \nn \\
    U_S(\vec b_\perp, \mu_0, \mu, a) &=
    \exp\Bigl[-\frac{4}{a} K_\Gamma \Bigr]\biggl(\frac{4}{\vec b_\perp^2 \mu_0^2 e^{2\ga_E}}\biggr)^{2\eta_\Ga/a}
    \,, \nn \\
    K_\Ga &= - \frac{\Ga_0}{4\beta_0^2} \biggl[ \frac{4\pi}{\alpha_s(\mu_0)} \Bigl(1 - \frac{1}{r} - \ln r\Bigr) + \Bigl(\frac{\Ga_1}{\Ga_0} - \frac{\beta_1}{\beta_0}\Bigr)(1-r+\ln r) + \frac{\beta_1}{2\beta_0} \ln^2 r \biggl]\,,
    \nn \\
      \eta_\Ga &= - \frac{\Ga_0}{2 \beta_0} \biggl[\ln r + \frac{\alpha_s(\mu_0)}{4\pi} \Bigl( \frac{\Ga_1}{\Ga_0} - \frac{\beta_1}{\beta_0}\Bigr) (r-1) \biggr]
\,.\end{align}
\end{widetext}
where $r = \alpha_s(\mu)/\alpha_s(\mu_0)$ and 
\begin{align} \label{eq:gabeta}
   \Ga_0 &= 4C_F\,, \nn \\
   \Ga_1 &= 4 C_F \biggl[\Bigl(\frac{67}{9}-\frac{\pi^2}{3}\Bigr) C_A - \frac{20}{9} T_F n_f\biggr]\,, \nn \\
   \beta_0 &= \frac{11}{3} C_A - \frac{4}{3} T_F n_f\,, \nn \\
   \beta_1 &= \frac{34}{3} C_A^2 - \Bigl(\frac{20}{3} C_A + 4 C_F\Bigr) T_F n_f
\,.\end{align}

\bibliography{vba.bib}

\end{document}